\documentclass[letterpaper, 10 pt, conference]{ieeeconf}  

\IEEEoverridecommandlockouts                              
\usepackage{cite}

\usepackage{amsmath,amssymb,amsfonts,amsthm}
\usepackage{graphicx}
\usepackage{textcomp}
\usepackage{xcolor}
\usepackage{caption}
\usepackage{subcaption}
\usepackage[draft]{todonotes}
\usepackage{algorithm}
\usepackage{algorithmic}
\DeclareMathOperator*{\argmax}{arg\,max}

\title{\LARGE \bf
Towards an Unified Structure for Reinforcement Learning: \\
an Optimization Approach
}

\author{Jicheng Shi$^\dagger$, Yingzhao Lian$^\dagger$, and Colin N. Jones
\thanks{$^\dagger$The first two authors contributed equally}
\thanks{This work has received support from the Swiss National Science Foundation under the RISK project (Risk Aware Data Driven Demand Response, grant number 200021 175627)}
\thanks{Jicheng Shi, Yingzhao Lian and Colin N. Jones are with Automatic Laboratory, Ecole Polytechnique Federale de Lausanne, Switzerland. 
        {\tt\small
        $\{$jicheng.shi, yingzhao.lian, colin.jones$\}$@epfl.ch}}%
}

\begin{document}

\maketitle
\thispagestyle{empty}
\pagestyle{empty}

\begin{abstract}
Both the optimal value function and the optimal policy can be used to model an optimal controller based on the duality established by the Bellman equation. Even with this duality, no parametric model has been able to output both policy and value function with a common parameter set. In this paper, a unified structure is proposed with a parametric optimization problem. The policy and the value function modelled by this structure share all parameters, which enables seamless switching among reinforcement learning algorithms while continuing to learn. The Q-learning and policy gradient based on the proposed structure is detailed. An actor-critic algorithm based on this structure, whose actor and critic are both modelled by the same parameters, is validated by both linear and nonlinear control. 
\end{abstract}

\section{Introduction}
Reinforcement learning (\textbf{RL}) learns an optimal control strategy by modelling a policy and/or a value function. In particular, given the current state, a policy returns control inputs, while a value returns the cost-to-go. Even though the optimal policy and the optimal value function show a strong duality with respect to the Bellman equation~\cite{bellman1966dynamic}, the usual parametric structures used in RL algorithms can either model the policy or the value function with a single parameter set. In view of the duality between the optimal policy and the optimal value function, some preceding works have tried to combine the policy model and the value function into a single model~\cite{lapan2018deep}, but their final structure can still only partially share information between the policy and the value function, such as common features learned from images with convolution layers~\cite{wang2015dueling,kaplan2017beating}. 

In this paper, we refer to a structure that outputs both the policy and the value function with a fully shared parameter set a \emph{unified structure}. Such a structure offers more flexibility in algorithmic development. For example, in a recommender system~\cite{zhang2019deep}, modelling a policy is preferable when the full information of a user is collected, while users with partial information available can only be modelled by a value function. With a unified structure, a continuous improvement of the control law can be achieved by seamlessly switching its learning algorithm based on the available information. Inspired by the Bellman equation, we propose to use a parametric optimization problem to achieve a unified structure, whose optimal value and optimal solution model value function and the underlying policy respectively. In this paper, the proposed unified structure is based on a parametric convex optimization problem, which has shown a strong connection to RL via nonlinear model predictive control~\cite{zanon2019safe,gros2019data,zanon2019practical}. In addition, parametric convex optimization problems are also used to model system dynamics~\cite{belanger2017end} and to reproduce expert's strategies~\cite{amos2018differentiable}. The contributions of this paper is summarized as follows:
\begin{itemize}
    \item Propose a unified structure that models both the policy and value function with a shared parameter set.
    \item Propose an amortized reparametrization method that adapts the proposed structure to policy gradient algorithms.
    \item A data efficient actor-critic algorithm whose actor also plays the role of the critic.
    \item Practical heuristics that enable more efficient and stable training.
\end{itemize}

\section{Preliminaries} \label{sect:premn}
In this section we will introduce the basic concepts of RL and two major RL algorithms: Q-learning and policy gradient.

\subsection{Reinforcement Learning}
Reinforcement learning considers a Markov Decision Process \textbf{(MDP)}, whose states $x\in\mathbb{R}^n$ forced by control inputs $u\in\mathbb{R}^m$ evolve under the dynamics $P(x^+|x,u)$. For the sake of clarity, we assume that the state is fully observable.

The reward for a trajectory $\tau =\left( x_0,u_0,x_1,u_1,... \right)$ is defined as
\begin{align*}
    R\left( \tau \right) =\sum_{t=0}^{\infty}{\gamma ^t}I(x_t,u_t), \gamma \in \left( 0,1 \right],
\end{align*}
where $\gamma$ and $I(x_t,u_t)$ denote the forgetting factor and the stage reward respectively. For the sake of clarity, we use $\gamma=1$ in the following. A control policy maps a state to a control input, $\pi:=x\rightarrow u$. Given a state $x$, each control policy $\pi$ is evaluated by its expected reward, called the \textbf{value function}
\begin{equation}\label{eqn:value_func}
 V^\pi(x)=\mathbb{E}_{\tau\sim\pi}(R(\tau)|x_0=x)).    
\end{equation}
  
A \textbf{Q-function} is defined as the expected reward for the policy $\pi$ with fixed state $x$ and the first control input $u$, $Q^\pi(x,u)=\mathbb{E}_{\tau\sim\pi}(R(\tau)|x_0=x,u_0=u)$. Correspondingly, the advantage of a control input $u$ at state $x$ with respect to its default policy $\pi$ is defined as:
\begin{equation}\label{eqn:advantage}
    A^\pi(x,u) = Q^\pi(x,u)-V^\pi(x)
\end{equation}

An optimal policy $\pi^*$ receives the highest reward in expectation and its optimality is characterized by the Bellman equation:
\begin{equation}\label{eqn:bellman_value}
    V^{\pi^*}(x_0) = \max\limits_u(I(x_0,u_0)+\mathbb{E}_{x_1}(V^{\pi^*}(x_1)))\;,
\end{equation}
and 
\begin{equation}\label{eqn:bellman_policy}
    \pi^*(x_0) = \argmax\limits_u(I(x_0,u_0)+V^{\pi^*}(x_1))\;.
\end{equation}

\subsection{Q-learning and Policy Gradient}
The policy and its corresponding value function are two main ingredients in all RL algorithms. In particular, Q-learning is the main algorithm used to improve the value function, while policy gradient is the base recipe for improving the policy directly. Specifically, Q-learning minimizes the temporal difference regarding the Bellman equation~\eqref{eqn:bellman_value}:
\begin{align*}
    T = I(x_i,u_i) + \max_{u_{i+1}}Q_\theta(x_{i+1},u_{i+1}) - Q_\theta(x_i,u_i)\;,
\end{align*}
where value function Q-functions is parametrized by some parameter $\theta$. Gradient descent is therefore used to update $\theta$.

Unlike Q-learning, a parametric stochastic policy $\pi_\theta$ with $u\sim\pi_\theta(x)$ is used in the policy gradient. The algorithm tilts the control input distribution towards the input sequences that experienced higher rewards $R(\tau)$. Specifically, a policy gradient algorithm maximizes the expected return of a given policy 
\begin{equation}\label{eqn:expected_reward}
J(\pi_\theta)=\underset{\tau\sim\pi_\theta}{\mathbb{E}}(R(\tau))\;. 
\end{equation}

The gradient of the expected return with respect to its parameter $\theta$ is evaluated as
\begin{equation} \label{eqn:policy_gradient}
    \nabla_\theta J(\pi_\theta)=\underset{\tau\sim\pi_\theta}{\mathbb{E}} (\sum\limits_{t=0}^T\nabla_\theta\log\pi_\theta(u_t|x_t)R(\tau))\;,
\end{equation}
with which gradient ascent is applied accordingly.

\section{Unified Reinforcement Learning Structure}\label{sect:uni}
The main challenge to a unified structure is the choice of the parametric model, where at least two problems need to be resolved:
\begin{itemize}
    \item Modelling a value function requires a scalar output while the output of a policy depends on the dimension of the control input $u$.
    \item A shared parametrization should reflect the relation between the policy and its underlying value function.
\end{itemize} 
We therefore propose to use a parametric optimization problem to model these two components, whose optimal solution can be mapped to the control inputs while the value function is modelled by the optimal value. In particular, this idea is consistent with the duality formed by the Bellman equation~\eqref{eqn:bellman_value} and~\eqref{eqn:bellman_policy}.

\subsection{Unified Structure with Lifting}
In principle, any parametric continuous optimization problem can be used to realize a unified structure. In this paper, we use a parametric quadratic programming (\textbf{QP}). Specifically, a parametric QP maps $z$ to the optimal solution $\mathcal{Q}(z)$ of the following problem:
\begin{align}\label{eqn:param_qp}
    \underset{\mu}{\max} &\frac{1}{2}\mu^TQ\mu + q^T\mu\\
\notag s.t.\; & Az+B\mu=b, Cz+D\mu \leqslant d\;,
\end{align}
where $\{Q,\;q,\;A,\;B,\;b,\;C,\;D,\;d\}$ are parameters of the problem with $Q$ negative-definite and the derivative of this QP is elaborated in Appendix~\ref{sect:cvx_layer}. Additionally, the corresponding function mapping $z$ to the optimal value is defined by $$\zeta(z):=\frac{1}{2}\mathcal{Q}(z)Q\mathcal{Q}(z)+q^T\mathcal{Q}(z)\;.$$

In general, $\mathcal{Q}(z)$ is used to model the policy while $\zeta(z)$ is used to model the value function. However, the concavity of this value function limits its applications. In order to accommodate nonlinear and more complex system dynamics, we propose to lift the problem with a parametric lifting function $f_\beta:x\rightarrow z$, where the lifting function is parametrized by $\beta$. In order to have higher flexibility, $\mu$ is assumed to lie within a higher dimensional space than $u$, the policy is then reconstructed by a fixed linear map $u=K\mathcal{Q}(z)$. Basically, any nonlinear parametric function can be used as a lifting function. In this paper, we use a multi-layer neural network to model the lifting $f_\beta$. With all the aforementioned components, the value function, the Q-function and the policy are defined accordingly as follows:
\begin{equation} \label{eqn:opt_value_func}
\begin{split} 
V_\theta(x)=\underset{\mu}{\max} &\frac{1}{2}\mu^TQ\mu + q^T\mu
\\
s.t. & Az+B\mu=b, \;Cz+D\mu \leqslant d,\\
&f_\beta(x)=z\;,
\end{split}
\end{equation}
\begin{equation} \label{eqn:opt_q_func}
\begin{split} 
Q_\theta(x,u)= \underset{\mu}{\max} &\frac{1}{2}\mu^TQ\mu + q^T\mu
\\
s.t. & Az+B\mu=b,\;Cz+Dz \leqslant d,\\
    & f_\beta(x)=z, K\mu=u\;,
\end{split}
\end{equation}
\begin{equation} \label{eqn:opt_policy}
    \begin{split} 
\pi_\theta(x)= K\underset{\mu}{\argmax} &\frac{1}{2}\mu^TQ\mu + q^T\mu
\\
s.t. & Az+B\mu=b, Cz+D\mu \leqslant d\\
    & f_\beta(x)=z\;.
\end{split}
\end{equation}
The parameters used for this parametric model are refered to as $\theta$
$$\theta:=\{Q,\;q,\;A,\;B,\;,b,\;C,\;D,\;d,\;\beta\}\;.$$
The RL algorithms that updates this structure will be elaborated in the next section.

\subsection{RL algorithms for the Unified Framework}
With Q-function and value function defined in Equations~\eqref{eqn:opt_value_func} and~\eqref{eqn:opt_q_func},  the  temporal difference can be rewritten as:
\begin{align*}
    T = &I(x_i,u_i) + \max_{u_{i+1}}Q(x_{i+1},u_{i+1})-Q_\theta(x_i,u_i)\;\\
    =& I(x_i,u_i)+V_\theta(x_{i+1})-Q_\theta(x_i,u_i)\;.
\end{align*}
The corresponding Q-learning algorithms can then be implemented accordingly.

Designing a policy gradient algorithm is more challenging, because the control law defined by the proposed structure is deterministic. A preceding work in~\cite{gros2019towards} proposes to perturb the parametric QP to enable a stochastic output, but this method is computationally intensive. Instead, we propose to perturb the optimal solution directly with an amortized reparametrization trick~\cite{kingma2013auto} as
\begin{equation} \label{eqn:amor_policy}
\begin{split}
    \tilde{\pi}_\theta(x) \sim& \enspace\mathcal{N}(\pi_\theta(x),\Sigma)\\
    \tilde{\pi}_\theta(x) =&\enspace \pi_\theta(x)+L\epsilon\\
    \Sigma = L^TL,&\enspace\epsilon\sim\mathcal{N}(0,I).
\end{split}
\end{equation}
The amortized reparametrization is widely used in variational inference~\cite{wainwright2008graphical}, it converts a deterministic control policy into a stochastic one while maintaining differentiability of the original problem. Meanwhile, the matrix $L$ is updated to enable an adaptive search. Note that apart from the Gaussian distribution in~\eqref{eqn:amor_policy}, most distributions from exponential family, such as gamma distribution, can be deployed to enable a better exploration~\cite{ruiz2016generalized}

\textbf{Remark 1}: The general framework of Q-leaning is seamlessly applicable in continuous action space. However, deep Q-leaning (\textbf{DQN}) is usually customized for environments with discrete action spaces \cite{mnih2013playing}, while the continuous action space is handled by other variants of Q-learning, such as Deep Deterministic Policy Gradient (\textbf{DDPG})~\cite{silver2014deterministic}.

\subsection{Conclusion}
We conclude this section by introducing one useful application of the proposed unified structure: an actor-critic RL algorithm. In principle, an actor-critic algorithm consists of an actor, which chooses an action given the current state, and a critic, which evaluates the performance of the actor. Specifically, the actor models the policy while the critic models the value function. Normally, these two components are parametrized independently with at most partially shared parameters. With the proposed unified framework, both the actor and the critic can be achieved by only one shared parameter set. During the training, improvement of the critic will also help to improve the performance of the actor, which is also true in the reverse. The data efficiency is therefore higher than standard structures, such as neural networks.

In Algorithm~\ref{alg:A2C}, an Advantage Actor Critic (A2C) algorithm with the proposed unified structure is detailed. Unlike general policy gradient in~\eqref{eqn:policy_gradient}, the gradient is instead calculated with respect to the advantage function $A(x_i,u_i)$ (Equation~\eqref{eqn:advantage}) to lower the variance~\cite{mnih2016asynchronous}, 
\begin{equation} 
\begin{split}
\nabla _{\theta}J\left( \pi _{\theta} \right) =\underset{\tau \sim \pi _{\theta}}{\text{E}}\left[ \sum_{t=0}^T{\nabla _{\theta}}\log \pi _{\theta}\left( x_t|u_t \right) A_\theta(x_i,u_i) \right]\;.
\end{split}
\end{equation}
It is noteworthy to point out that the value function is updated with gradient descent, while the policy is updated with gradient ascent. Note that the parameters of the critic can also be updated at a lower frequency and with a soft gradient~\cite{haarnoja2018soft}, similar to the idea of target networks in~\cite{mnih2013playing}:
\begin{equation}
\begin{split}
    & \hat{\theta} \leftarrow \theta + \alpha_c A_\theta\nabla_\theta V_\theta \\
    & \theta \leftarrow \tau \hat{\theta} + (1-\tau)\theta\;.
\end{split}
\end{equation}

In conclusion, the proposed unified framework enables a seamless switch among many RL algorithms without interrupting training. It also shows higher data efficiency when applied to actor-critic algorithms. It is noteworthy that it can also be deployed in other algorithms such as DDPG \cite{lillicrap2015continuous}, and Soft Actor Critic (SAC)\cite{haarnoja2018soft}.

\begin{algorithm}[h] 
\caption{A2C with the unified framework}
\begin{algorithmic}[1] \label{alg:A2C}
\STATE Initialize convex optimization layer for value function, and policy function with parameters $\theta$
\STATE Initialize step counter $t=1$
\FOR{episode $\,=\,1,2,...,M\,$}
    \STATE Reset the gradients in "actor" and "critic" part: $d\theta = 0$ and $d\theta = 0$
    \STATE Reset the environment, $done = FALSE$, $t_{start} = t$
    \WHILE{not $done$ or $t-t_{start}>t_{max}$}
        \STATE Observe state $x_t$, execute action $u_t$ according to the policy, $\pi_{\theta} (x_t) $ of by~\eqref{eqn:amor_policy}
        \STATE Observe next state $x_{t+1}$, stage cost $I_t$, and $done$ signal
        \STATE $t\gets t+1$
    \ENDWHILE
    \STATE $R = \left\{ \begin{array}{ll}
	0 & \text{for terminal} x_t\\
	V_{\theta}\left( x_t \right) & \text{for non-terminal } x_t\\
\end{array} \right. $
    \FOR{i=$\,\,t-1,t-2,...,0 \,$}
        \STATE  $R \leftarrow I_i + R$
        \STATE Accumulate gradients in "actor" part, $d \theta \leftarrow d \theta +\alpha_a \nabla_{\theta} \log \pi\left(u_{i} | x_{i} \right)A(x_i,u_i)$
        \STATE Accumulate gradients in "critic" part: $d \theta \leftarrow d \theta -\alpha_c\partial\left(R-V_{\theta }\left(x_{i}\right)\right)^{2} / \partial \theta$
    \ENDFOR
    \STATE Update $\theta$ as: $\theta\leftarrow\theta+d\theta$.
\ENDFOR
\end{algorithmic}
\end{algorithm}

\section{Practical Issues in Training} \label{sect:prac}
In this section, we will talk about a few issues that enable a better training performance.
\subsection{Lifting Selection} \label{sect:lift_selec}
The lifting $f_\beta$ encodes more information about the states by mapping it to a lifting space. Without a proper initialization of the parameter $\beta$, the training will be difficult because it learns both the optimization problem and the lifting together. As suggested in~\cite{zanon2019practical}, the lifting function can be perceived as a model of the system, which implies that we can use system identification to initialize the parameters in this model. For example, $f_\beta$ for a nonlinear system can be initialized by the lifting function of a Koopman operator~\cite{williams2015data,koopman1932dynamical}, where the unsupervised learning algorithm in~\cite{lian2019learning} can be applied. Note that the lifting space of $z$ in this case is of higher dimension than the state space.

State $x$ is not necessarily available in practice, instead, the measurements from sensors with redundant information are used in RL algorithms, such as video stream used in autonomous vehicle. In this case, mapping high dimensional measurements to an informative low dimensional features can improve computational efficiency. Based on the idea of transfer learning~\cite{pan2009survey}, one can initialize the lifting with some pretrained neural networks, such as ResNet~\cite{he2016deep}.

\subsection{Miscellaneous}
The parametric optimization problem is required to be feasible during the training. Without the a priori knowledge of the feasible set, a soft-constrained problem can be applied to ensure feasibility,
\begin{equation}\label{eqn:uni_frame_soft_constrained}
\begin{split} 
\underset{\mu}{\max} &\frac{1}{2}\mu^TQ\mu + q^T\mu
+ \rho(\varepsilon)\\
s.t. & Az+B\mu=b, Cz+D\mu \leqslant d+\varepsilon
\end{split},
\end{equation}
where $\varepsilon$ denotes the slack variables with $\rho(\varepsilon)$ is the penalty for the constraint violation. 

\section{EXPERIMENTS} \label{sect:exp}
In this section, both linear and nonlinear systems are used to validate the proposed structure with Q-learning and the actor-critic Algorithm~\ref{alg:A2C}. The proposed structure and its automatic differentiation are implemented with \texttt{CVXPylayers}~\cite{agrawal2019differentiable} and \texttt{Pytorch}~\cite{paszke2017automatic}, while the validation simulation is carried out on the \textsc{OpenAI Gym} platform~\cite{brockman2016openai}. The policy gradient is implemented but not reported for the sake of compactness.

\subsection{Linear System}
We consider a controllable but unstable linear system with box constraints, 
\begin{equation} 
\begin{split}
x^+=&\left[ \begin{array}{rr}
	1.53 & 0.25\\
	-0.56 &-0.52\\
\end{array} \right] x+\left[ \begin{array}{r}
	1.23\\
	-0.96\\
\end{array} \right] u
\\
&x\in[-4,\;4],\;u\in[-1,\;1],\;
\end{split}
\end{equation}
whose stage reward is given by $I(x,u)=-(\lVert x \rVert_2^2 +\lVert u \rVert_2^2)$. Each run of the simulation is initialized with uniform samples from $[-2,\;2]$ and terminated when $x\in[-0.1,\;0.1]$ or $x\not\in[-4,\;4]$.

A rough system identification with only $20$ measurements is applied to initialize the linear lifting function: 
\begin{equation} 
\begin{split}
z =\left[ \begin{array}{c}
    	I\\
	A\\
	...\\
	A^9\\
\end{array} \right] x+\left[ \begin{array}{lllll}
	0 & 0 &... &... & 0 \\
	B & 0 &... &... & 0               \\
	AB &B &0 &... &0             \\
	... &... &...&...&...\\
	A^{8}B &A^{7}B &...&... &B\\
\end{array} \right] \mu
\end{split}\;,
\end{equation}
where $A,\;B$ are identified from linear system $x^+=Ax+Bu$. The reconstruction $K$ is chosen to be $[1,0,\dots,0]$ and other parameters are initialized randomly.

\begin{figure}[htb!]
    \centering
    \begin{subfigure}[t]{0.5\textwidth}
        \includegraphics[height=5cm]{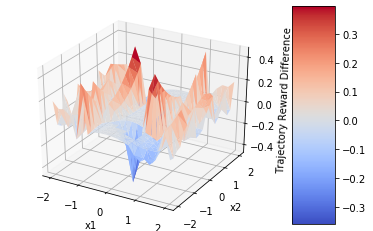}\\
        \caption{Q-learning}
    \end{subfigure}\\
    \begin{subfigure}[t]{0.5\textwidth}
        \includegraphics[height=5cm]{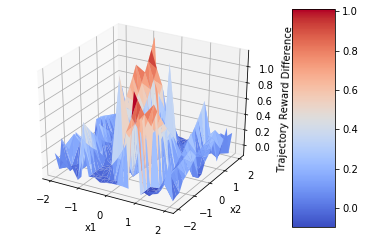}\\
        \caption{Actor-Critic}
    \end{subfigure}
        \caption{\label{fig:compare_MPC}Comparison of trajectory reward against the MPC controller}
\end{figure}

The control law learnt by the actor critic algorithm is compared against a stable model predictive controller~\cite{camacho2013model}, whose prediction horizon is $10$. The correct model is used in the MPC controller, whose performance is sub-optimal against the infinite-horizon solution but has good balance between performance and prediction horizon. The cumulative reward (Equation~\eqref{eqn:expected_reward}) of the learnt policy and the MPC controller is estimated by the Monte-Carlo method, and the learned policy achieves 2.07\% higher reward than the MPC controller. The same procedure is carried out for Q-learing with the proposed structure, which also receives 0.24\% higher expected reward than MPC. Figure~\ref{fig:compare_MPC} shows the corresponding comparisons, where the actor-critic algorithm achieves a higher reward regardless of the initial state while the Q-learning has a lower reward for some initial states.

\begin{figure}[htb!]
    \centering
    \begin{subfigure}[t]{0.5\textwidth}
        \includegraphics[height=5cm]{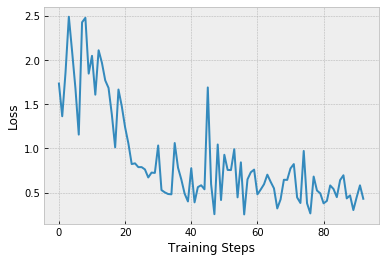}
        \caption{Q-learning}
    \end{subfigure}\\
    \begin{subfigure}[t]{0.5\textwidth}
        \includegraphics[height=5cm]{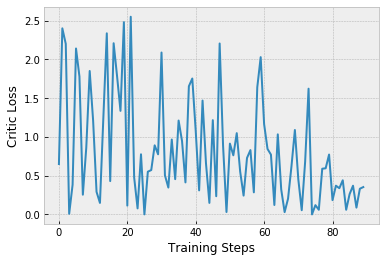}
        \caption{Critic loss in actor-critic algorithm}
    \end{subfigure}
        \caption{\label{fig:convergence_linear}Convergence in the RL algorithm}
\end{figure}

Even though the actor-critic algorithm shows better performance in terms of the final learned policy, its convergence is somewhat less stable (Figure~\ref{fig:convergence_linear}). This is caused by the update of both the actor and the critic on a shared parametrization. However, both algorithms converge within 100 iterations, while the conventional RL algorithms based on deep neural networks require more than $10^4$ iterations~\cite{van2016deep}, which shows that the proposed structure has significantly higher data efficiency for this toy example.

\subsection{Nonlinear System}
\begin{figure}[htb!] 
    \centering
        \includegraphics[height=5cm]{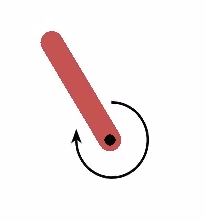}
        \includegraphics[height=5cm]{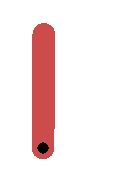}
        \caption{\label{fig:pen}Screen shots from the pendulum simulation with arrow indicating the torque~\cite{brockman2016openai}}
\end{figure}

We consider an inverted pendulum~(Figure~\ref{fig:pen}), whose dynamics are given by
\begin{equation} 
\begin{split}
&\theta^{+}=\theta + \Dot{\theta} \cdot dt \;,\\
    &\Dot{\theta}^{+}= \Dot{\theta} - \frac{3g}{2l} \cdot dt \cdot \sin{\theta+\pi} + \frac{3}{2ml} \cdot dt \cdot u \;,
\end{split}
\end{equation}
where the constant parameters are $g=10,m=1,l=1$ and $dt=0.05$. In addition, the state constraints and input constraints are $\Dot{\theta}\in[-8,\;8]$ and $u\in[-10,\;10]$. The stage reward is chosen to be $I(x,u)=-\theta^2 -0.1\Dot{\theta}^2-0.001u^2$.

The lifting function $f_{\beta}$ is initialized by the Koopman operator based on a linearly-recurrent autoencoder network~\cite{otto2019linearly}. Then the two-dimension state $x=\left[\theta,\Dot{\theta}\right]$ is mapped to a five-dimension $z$. Other parameters are initialized randomly.

\begin{figure}[htb!]
    \centering
    \begin{subfigure}[t]{0.5\textwidth}
        \includegraphics[height=5cm]{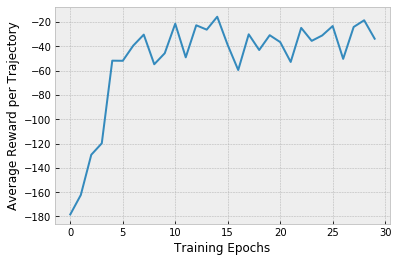}
        \caption{Q-learning}
    \end{subfigure}\\
    \begin{subfigure}[t]{0.5\textwidth}
        \includegraphics[height=5cm]{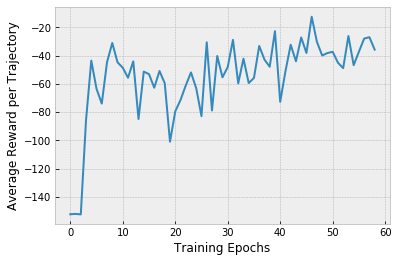}
        \caption{Critic loss in actor-critic algorithm}
    \end{subfigure}
        \caption{\label{fig:convergence_reward}Improvement of the expected rewards}
\end{figure}

Both the Q-learning and AC based on our proposed framework are tested, and the Monte-Carlo estimated expected return during training are shown in Figure~\ref{fig:convergence_reward}. Both algorithms converges within 50 iterations, which also demonstrate a higher data efficiency of the proposed structure.

\begin{figure}[htb!]
    \centering
        \includegraphics[height=6.3cm]{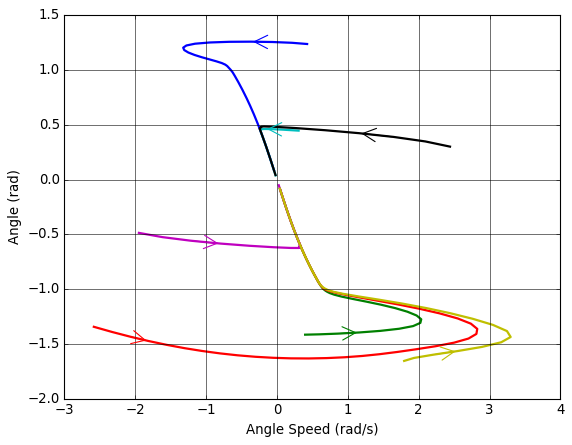}
    \caption{\label{fig:A2C_traj}Trajectories with random initial states under actor-critic control policy}
\end{figure}

With the aforementioned two RL policies, the pendulum quickly moves to the target position and stays still. Trajectories with random initial states under actor-critic control policy are shown in Figure~\ref{fig:A2C_traj}.

\section{CONCLUSION} \label{sect:conc}
In this paper, a unified structure based on a parametric optimization problem is proposed, whose Q-learning and actor-critic algorithms are correspondingly derived. The proposed unified structure has a higher flexibility and results in a data-efficient actor-critic algorithm. 


\bibliographystyle{./bibliography/IEEEtran}
\bibliography{mylib}

\section{Appendix}\label{sect:app}
\subsection{Derivative of a Parametric Convex Optimization Problem}\label{sect:cvx_layer}
The existence of the Fréchet derivative of a parametric convex optimization is guaranteed by its uniform level boundeness~(Theorem 1.17 in~\cite{rockafellar2009variational}). Once the optimal solution is determined, the derivative with respect to the parameters are given by its KKT system~\cite{bonnans2013perturbation}.

For the sake of clarity, we elaborate this derivative by a standard quadratic programming (QP) form; please refer to~\cite{agrawal2019differentiable} for a general formulation. Consider the a parametric QP, $\mathcal{Q}(f):=f\rightarrow z^*$ with training parameters $\{Q,\;q,\;A,\;b,\;E\}$ and $Q$ negative definite:
\begin{equation}
\begin{split} \label{eqn:quaratic_program}
\underset{z}{\max}\ &\frac{1}{2}z^TQz + q^Tz
\\
\text{s.t.}\ & Az \leqslant b, Ez=f
\end{split}
\end{equation}
The KKT conditions for the QP are:
\begin{equation}
\begin{split} \label{eqn:KKT_conditon}
    Qz^* + q + A^T\lambda^* + E^T\nu^*  &= 0
    \\
    D\left( \lambda^* \right) (Az^*-b) &= 0
    \\
    Ez^* - f & = 0
\end{split}
\end{equation}
where $z^*, \nu^*, \lambda^*$ are the optimal primal and dual variables, $D(\cdot)$ builds a diagonal matrix from a vector. Then the differentials of KKT conditions can be computed as:
\begin{equation} \label{eqn:KKT_diff}
\begin{split}
    \left[\begin{array}{ccc}
        Q & A^T & E^T  \\
        D(\lambda^*)A & D(Az^*-b) & 0 \\
        E & 0 & 0
    \end{array} \right]
    \left[ \begin{array}{c}
         dz \\
         d\lambda \\
         d\nu
    \end{array}\right] \\
    = -\left[ \begin{array}{c}
         dQz^* + dq + dA\lambda^* + dE^T\nu^* \\
         D(\lambda^*)dAz^* - D(\lambda^*)db \\
         dEz^* - df
    \end{array}\right]
\end{split}
\end{equation}
The derivatives of $z^*$ with respect to the parameters ($Q,q,A,b,E$) and the function input $f$ are given by the solution to the linear system defined in Equation~\eqref{eqn:KKT_diff}. For example, the solution $dz$ of~\eqref{eqn:KKT_diff} gives the result of $\frac{\partial{z^*}}{\partial{Q}}$ if we set $dQ=I$ and the differentials of other parameters to 0. The gradient of optimal value $L(z^*)$ with respect to $Q$ is calculated accordingly  as $\frac{\partial{L(z^*)}}{\partial{z^*}}\frac{\partial{z^*}}{\partial{Q}}$.

\end{document}